\documentclass[12pt]{iopart}
\usepackage{graphicx}
\newcommand{\prl}{{\it Phys.\ Rev.\ Lett.} }
\newcommand{\prb}{{\it Phys.\ Rev.} B }

\begin{document}
\title{Molecular rolling friction: the cogwheel model}
\author{O M Braun$^{1,2}$ and Erio Tosatti$^{2,3,4}$} 
\address{$^{1}$ Institute of Physics, National Academy of Sciences of Ukraine, 03028 Kiev, Ukraine}
\address{$^{2}$ International School for Advanced Studies (SISSA), Via Beirut 2-4, I-34014 Trieste, Italy}
\address{$^{3}$ CNR-INFM Democritos National Simulation Center, Via Beirut 2-4, I-34014 Trieste, Italy}
\address{$^{4}$ International Centre for Theoretical Physics (ICTP), P.O. Box 586, I-34014 Trieste, Italy}
\ead{obraun@iop.kiev.ua (O.M.Braun)}

\begin{abstract}
With the help of a two-dimensional model we study rolling lubrication
by circular (``2D fullerenes'') molecules for a wide range of parameters.
The conditions under which microscopic rolling friction may be effective
are identified, and related to the relative ingraining between substrate and
molecule, the latter behaving as a nanosized cogwheel.
\end{abstract}

\pacs{68.35.Af, 81.40.Pq, 61.48.+c, 68.60.-p}

Keywords: molecular rolling friction; fullerenes

\submitto{J.\ Phys.: Condens.\ Matter} \bigskip
\section{Introduction}
\label{intro}

Feynman \cite{F1960} famously foreshadowed atomic-scale
machines that could perform similarly to their macroscopic analogs. Nowadays
the problem of designing nanomechanical devices, in particular, to reduce friction
by means of nano- and micro-bearings \cite{D1992} is real. The possible use of fullerene  C$_{60}$
for molecular nanobearings gave rise to Molecular Dynamics (MD) modeling
\cite{LGG2004,KH2004,SITM2007,KHFBHK2005}.
Other fullerene-like metal dichalcogenide MX$_2$ (where M=Mo or W and X=S or Se)
molecules were considered as additives in oil lubricants, and predicted to provide
interesting tribological properties (e.g., see \cite{GHET2004} and references therein).
Simulations showed that ball-shaped molecules may either slide or rotate
over a surface, depending on the substrate and the position of the molecule.
For example, C$_{60}$ slides on graphite in the AB configuration
(hexagonal C$_{60}$ ring lying flat on graphite), but rotates in the
on-top frustrated AB configuration
where one C$_{60}$ corner atom faces the center of the graphite hexagon.
The rolling configuration is characterized by very low friction \cite{LGG2004,SITM2007},
with a predicted friction coefficient of the order $\mu \sim 0.01-0.02$
\cite{KH2004} or even smaller \cite{SITM2007}.

Attempts to realize these ideas experimentally had
only limited success so far. A single C$_{60}$ molecule confined between two solid substrates
may begin to roll when a torque of order 
$10^{-19}$~Nm is applied~\cite{MKS2003}.
However, C$_{60}$ molecules actually condense to form close-packed layers,
as found, e.g., on a graphite substrate \cite{2345,78910}.
A single C$_{60}$ monolayer (ML) takes a crystalline structure with
2D spatial order at low temperatures. It undergoes a first-order orientational
order-disorder transition at $T = T_m \approx 260$ K~\cite{141516}, the molecules
exhibiting free rotation at $T > T_m$.
At $T_m$ there is an abrupt change in friction~\cite{LTXLX2003}, but
the lowest friction coefficient is of order $\mu \sim 0.15$
\cite{78910,LTXLX2003}, worse than with traditional oil-based lubricants.
Coffey and Krim \cite{CK2006} reported a quartz crystal microbalance study
of one or two C$_{60}$ monolayers adsorbed on Ag(111) or Cu(111).
There are no rotations in a C$_{60}$ ML on Cu(111), and only a slow change
of molecular orientations in the C$_{60}$/Ag(111) ML. For two ML instead,
C$_{60}$ molecules in the second layer rotate freely at 300 K.
However a molecularly thin methanol film deposited over the C$_{60}$,
failed to show either the expected low friction, or any essential difference
between these systems. Thus this particular nanobearing design apparently would not work.

Some charge transfer and bonding between C$_{60}$ and the metal substrate
may be held responsible for hindering the rolling. Another reason 
lies in the full layer coverage. Balls in macroscopic bearings are arranged
so as to prevent contact, but rolling molecules in the ML are always in
contact, hindering their mutual rolling and jamming
the same way two ingrained rolling cogwheels would.
As discussed earlier on \cite{B2005}, a way to
avoid jamming is to lower dramatically the coverage, well below one ML
(the molecule density should anyway be sufficient to prevent the two
surfaces from touching). In view of that, and in the lack of
well defined low coverage experiments, a study of the
single molecule rolling friction represents a natural starting point, and
indeed a revealing one.

Macroscopically, the main source of rolling friction of a ball or tire
comes from deformation. Both substrate and roller are (elastically or
plastically) deformed at the contact. The deformation energy is partly
released and lost as bulk frictional heat when the roller moves on \cite{P0}.
By designing the bulk so that dissipation is poor, rolling friction
can be made $10^2$ to $10^3$ times lower than the sliding friction;
the latter being due to adhesion, i.e., breaking and re-forming of slider-substrate bonds.
It the previous work~\cite{B2005} we studied molecular rolling friction for the system,
where the lubricant and both substrates were constructed of the same molecules,
so that the lubricant and substrates were deformable and commensurate.
The minimal friction coefficient found in simulation, was of order $\mu > 0.15$
in agreement with available experimental data.
Naturally it emerges a question, what would be a value of $\mu$
for the rigid substrates, when the losses due to deformation are absent, 
or for the case of incommensurate lubricant/substrate interface.
As the roller size is decreased however, adhesion grows in importance,
eventually becoming the main source of friction. 
To rotate a molecule, one has to break the molecule-substrate bonds
from one side of the molecule and create new bonds on the opposite side.
Thus, there are no reasons to expect that molecular rolling friction should be
much lower than the sliding friction.

\textit{Our present goal is to understand what could be the lowest friction coefficient
attainable for molecular rolling and which system parameters might provide it.}
Besides, we 
show that a macro- to microworld mapping does work,
but one has to choose properly the macroscopic counterpart, which in the
present case is a cogwheel.
Because we are interested in general trends, we 
explore a minimal two-dimensional (2D) model,
which allows us to span a large number of parameters, and also provides an easier
visualization of the processes inside the lubricant.

\section{Model}
\label{model}

We consider two substrates with lubricant molecules in between, all of them made
up of classical point particles (atoms). Atoms can move in the $(x,y)$ plane,
where $x$ is the sliding direction and $y$ is perpendicular to the substrates.
The substrates, pressed together by a load force $F_l = N_s f_l$,
consist of rigid atomic chains of length $N_s$ and equal lattice constant
$R_s$, so that the system size in the sliding direction is $L_x = N_s R_s$ and
the total mass of the substrate is $N_s m_s$ (we use periodic boundary condition
along $x$). The bottom rigid substrate is fixed at $x=y=0$,
the top one is free to move in both $x$ and $y$.
The top substrate is driven along $x$ with speed $v_s$
through a spring of elastic constant $k_s$.
The spring force $F$, whose maximum value before motion measures the static
friction force $F_s$, and whose average during smooth motion $F_k = \langle F \rangle$
is the kinetic friction force, is monitored
during simulation (throughout the paper we normalize forces per
substrate atom $f=F/N_s$). Thus, our model is a 2D variant of a typical
experimental setup in tribology \cite{P0,BN2006}.
Between the substrates we have circular (``spherical'') lubricant molecules
built as in Ref.~\cite{B2005}.
Each molecule 
has one central atom and $L$ atoms on circle of radius
$R_m = R_{ll} /2 \sin(\pi/L)$
so that their chord distance is $R_{ll}$.
They are coupled with the central atom,
additionally to the 12-6 Lennard-Jones (LJ) potential,
by stiff springs of elastic constant $K_m$,
$V_{\rm stab} (r)={1\over2} K_m (r-R_{\rm stab})^2$,
where the distance $R_{\rm stab}
 = R_m + (12 \, V_{ll} /K_m R_m) \left[ (R_{ll} /R_m)^{6} - (R_{ll} /R_m)^{12} \right]$
is chosen so that the total potential
$V_{\rm LJ} (r) + V_{\rm stab} (r)$ is minimum at $r=R_m$.
With $K_m =100$ the resulting stiff molecular shape
resisted destruction during the simulations.
All atoms interact via the LJ potential
$V_{\rm LJ} (r)=V_{\alpha \alpha^{\prime}} \left[ \left( {R_{\alpha \alpha^{\prime}}}/{r}
\right)^{12} -2 \left( {R_{\alpha \alpha^{\prime}}}/{r} \right)^{6} \right]$,
where $\alpha,\alpha^{\prime} = s$ or~$l$ for the substrate or lubricant atoms respectively.
Thus, the lubricant-lubricant interaction is described by the parameters
$V_{ll}$ and $R_{ll}$, while the lubricant-substrate interaction, by
$V_{sl}$ and $R_{sl}$
(direct interaction between the top and bottom substrates is omitted, as
they are not allowed to touch).
We use dimensionless units, where $m_s = m_l =1$,
$R_{ll} =1$, and the energy parameters $V_{\alpha \alpha^{\prime}}$
takes values around $V_{\alpha \alpha^{\prime}} \sim 1$.

Because a 2D model cannot reproduce even qualitatively the phonon spectrum of a 3D system,
and because frictional kinetics is generally diffusional rather than inertial,
we use Langevin equations of motion with Gaussian random forces
corresponding to temperature $T$, and a damping force
$f_{\eta, x} = -m \, \eta (y) \, \dot{x} -m \, \eta (Y-y) \, (\dot{x} - \dot{X})$,
where $x,y$ are the atomic coordinates and
$X,Y$ are the coordinates of the top substrate
(the force $f_{\eta, y}$ is defined in the same way).
The viscous damping coefficient is assumed to decrease with the distance
from the corresponding substrate,
$ \eta(y) = \eta_0 \left[ 1-\tanh (y/y_d) \right]$,
where typically $\eta_0 =1$ and $y_d \sim 1$.

We present simulation results for molecule friction
from $L=5$ (the simplest circular molecule) up to $L=13$ and~14,
which may be considered as a 2D version of fullerenes.
In fact in the 3D case, the surface area of the spherical molecule is $s=4\pi R_m^2$.
If we put $L_3 =60$ atoms on the surface, this gives $s \approx L_3 R_{ll}^2$,
or $R_m /R_{ll} \approx 2.18$. In 2D, the length of the circle is $2\pi R_m \approx L R_{ll}$,
or $L \approx 2\pi R_m /R_{ll}$, which leads to $L \approx 13.7$
for the same ratio $R_m /R_{ll}$ as for 3D fullerenes.

\section{Rigid molecule}
\label{rigid}

We first consider a rigid circular molecule, i.e.,
$V_{ll} = \infty$ and $K_m = \infty$. Let us fix $X$ of
the top substrate and seek minimum of the potential energy $V$ by varying
the coordinate $Y$ of the top substrate, and
the center $(x_c, y_c)$  and the rotation angle $\phi$ of the molecule.
The $X$ dependence of $V$, $(x_c, y_c)$, and $\phi$ defines the adiabatic
trajectory, which describes the joint substrate and lubricant motion when infinitely slow.
We define the activation energy $E_a = \max \left[ V(X) \right] - \min \left[ V(X) \right]$,
and the magnitude of the static friction force, approximated  as
$f_s = \max \left[ dV(X)/dX \right]$ ($f_s \sim E_a$ in our units).

\begin{figure}[h] 
\begin{center}
\includegraphics[width=12cm]{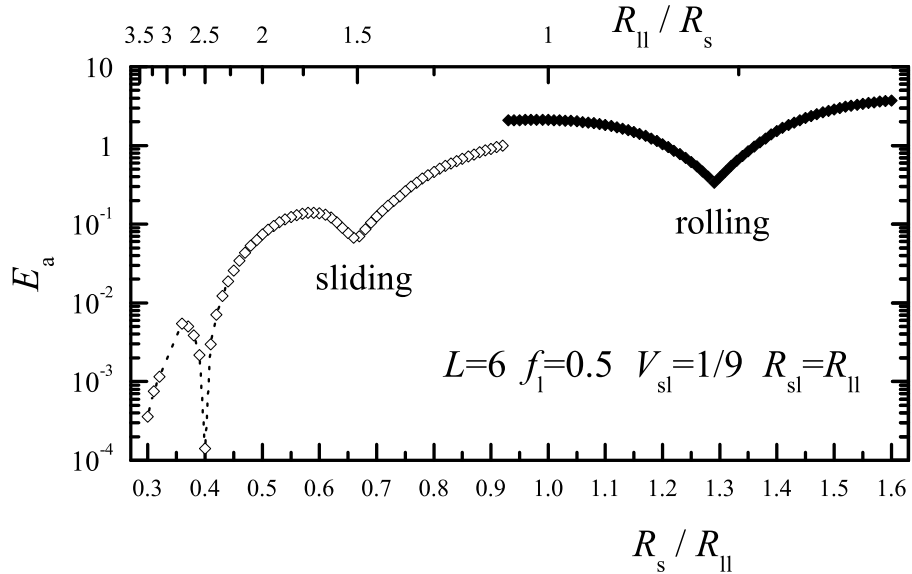} 
\end{center}
\caption{\label{R04c} 
Activation energy $E_a$ as a function of the ratio of the substrate lattice constant $R_s$
to $R_{ll}$
for the rigid $L=6$ molecule, for $f_l=0.5$, $V_{sl}=1/9$, and $R_{sl}=R_{ll}$.
Open symbols correspond to sliding, solid symbols to rolling.}
\end{figure}

\begin{figure}[t] 
\begin{center}
\includegraphics[clip,width=7cm]{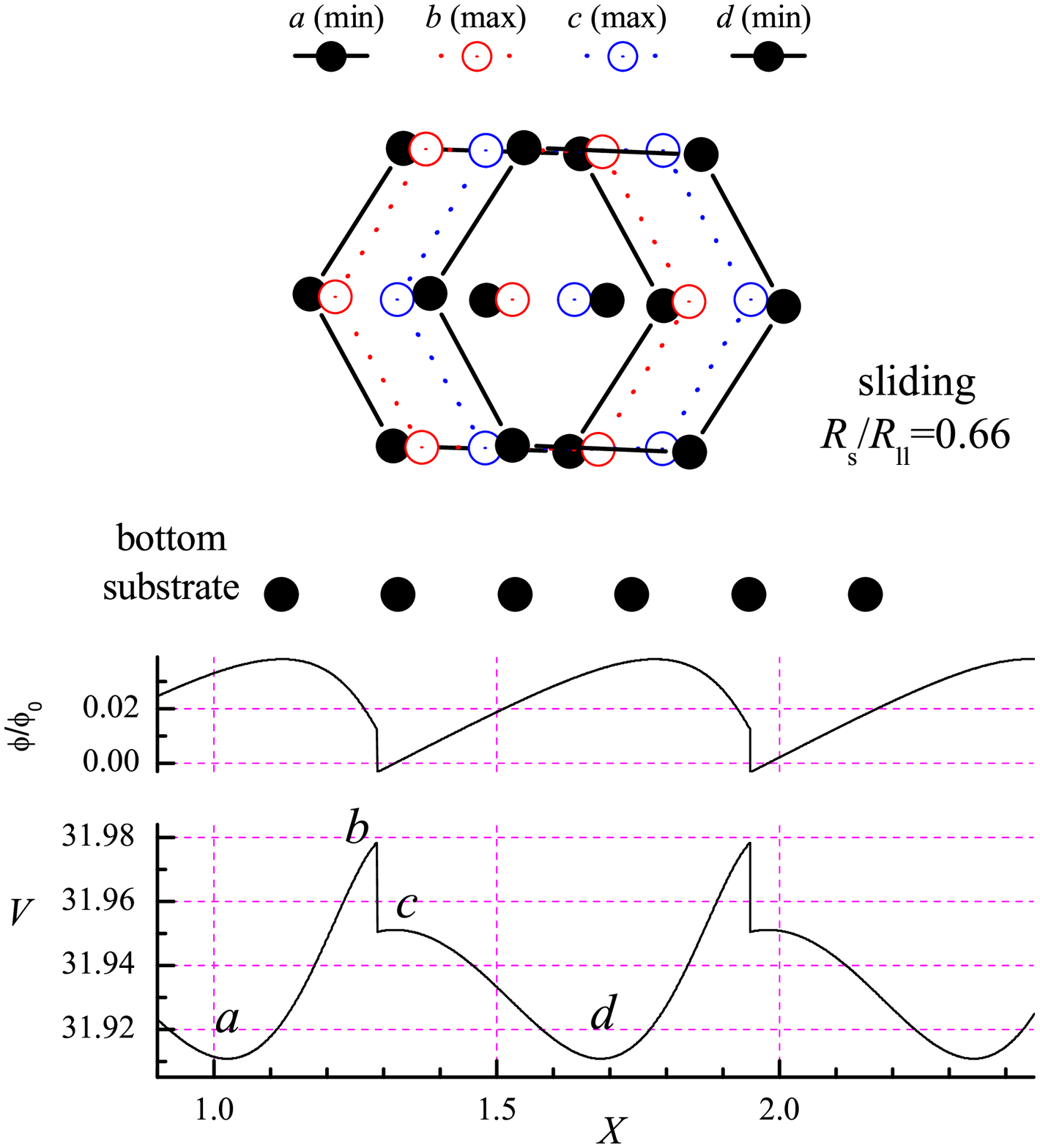} 
\hspace{1cm}
\includegraphics[clip,width=7cm]{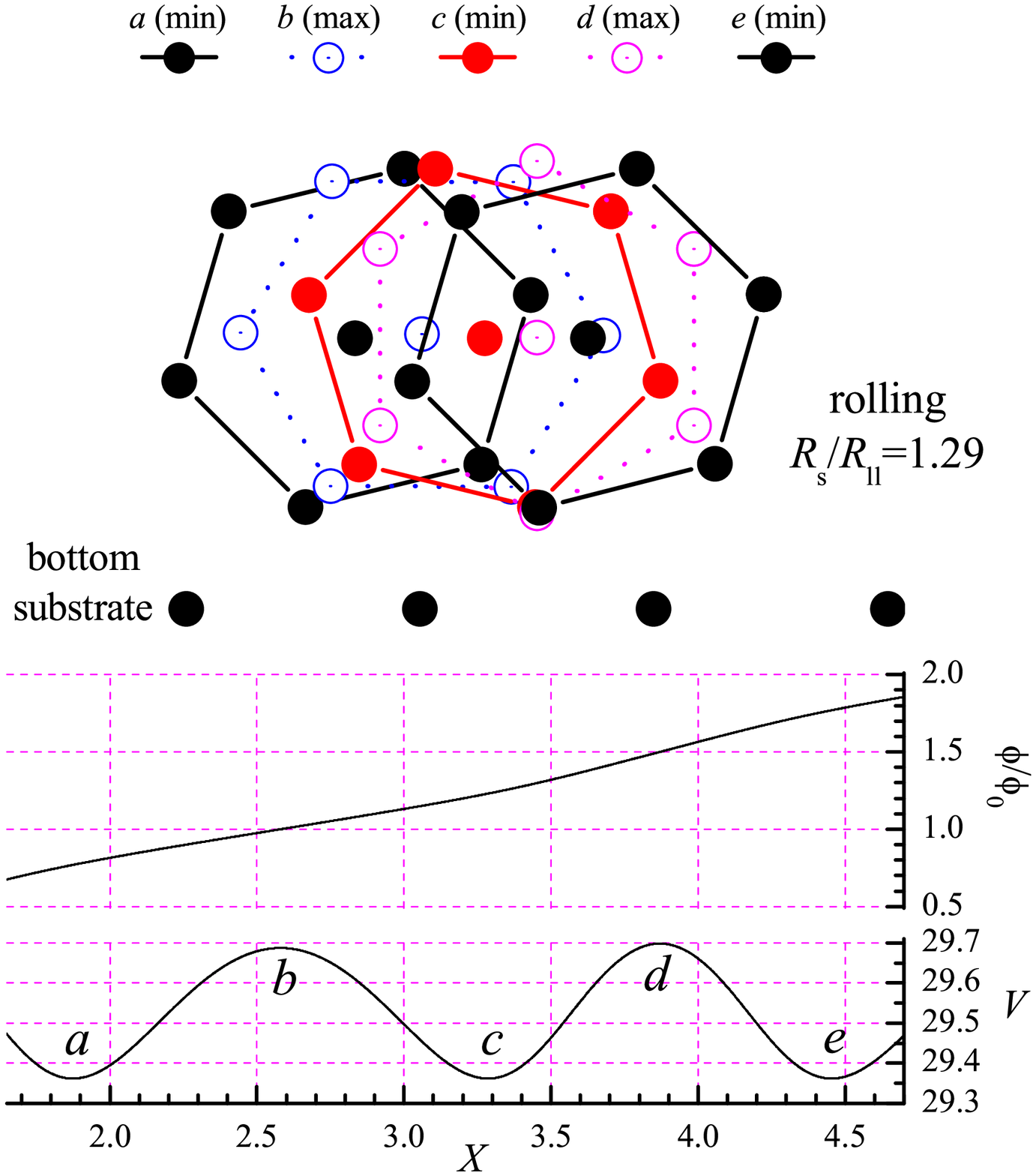} 
\end{center}
\caption{\label{R05} 
Sliding adiabatic motion of the rigid $L=6$ molecule for $R_s/R_{ll}=0.66$
($\Delta \phi < \phi_0$, left panel)
and rolling for $R_s/R_{ll}=1.29$ ($\Delta \phi > \phi_0$, right panel).
Other parameters as in figure~\ref{R04c}.
Lower panels: $X$-dependence of potential energy $V(X)$ and the rotation
angle $\phi (X)/\phi_0$, where $\phi_0 = 2\pi/L$. Top panel: configurations as the
molecule moves from one minimum of $V(X)$ to the next.}
\end{figure}

Figures \ref{R04c} and~\ref{R05} show the results for the $L=6$ molecule
when, to simplify further, $R_{sl}$ is kept constant, $R_{sl}=R_{ll}$.
The energy $V(X)$ is periodic with $R_s$ (or a multiplier of $R_s$).
The molecular angle $\phi$ varies by $\Delta \phi$
as the potential energy $V(X)$ changes from minimum to maximum.
Because $\phi (X)$ must be continuous,
the motion corresponds to sliding if $\Delta \phi < \phi_0 \equiv 2\pi/L$,
while if $\Delta \phi > \phi_0$ the molecule must rotate when it moves.
As figure~\ref{R04c} shows,
for $R_s < R_{ll}$ the motion corresponds to sliding, i.e.,
the molecule is shifted as a whole, slightly oscillating during motion
(figure~\ref{R05}, left panel).
Similarly to the motion of a dimer in a periodic potential \cite{B1990},
the activation energy has maxima at $R_{ll}=nR_{s}$ (where $n$ is an integer)
and minima at $R_{ll}=(n-1/2)R_{s}$.
On the other hand, for $R_{ll} < R_{s}$ the motion corresponds to rolling
(figure~\ref{R05}, right panel). Here $E_a(R_s)$ has minima at some values
of the ratio $R_{s}/R_{ll}$ (e.g., for $R_{s}/R_{ll} \approx 1.29$ in figure~\ref{R04c}).

\begin{figure}[h] 
\begin{center}
\includegraphics[width=10cm]{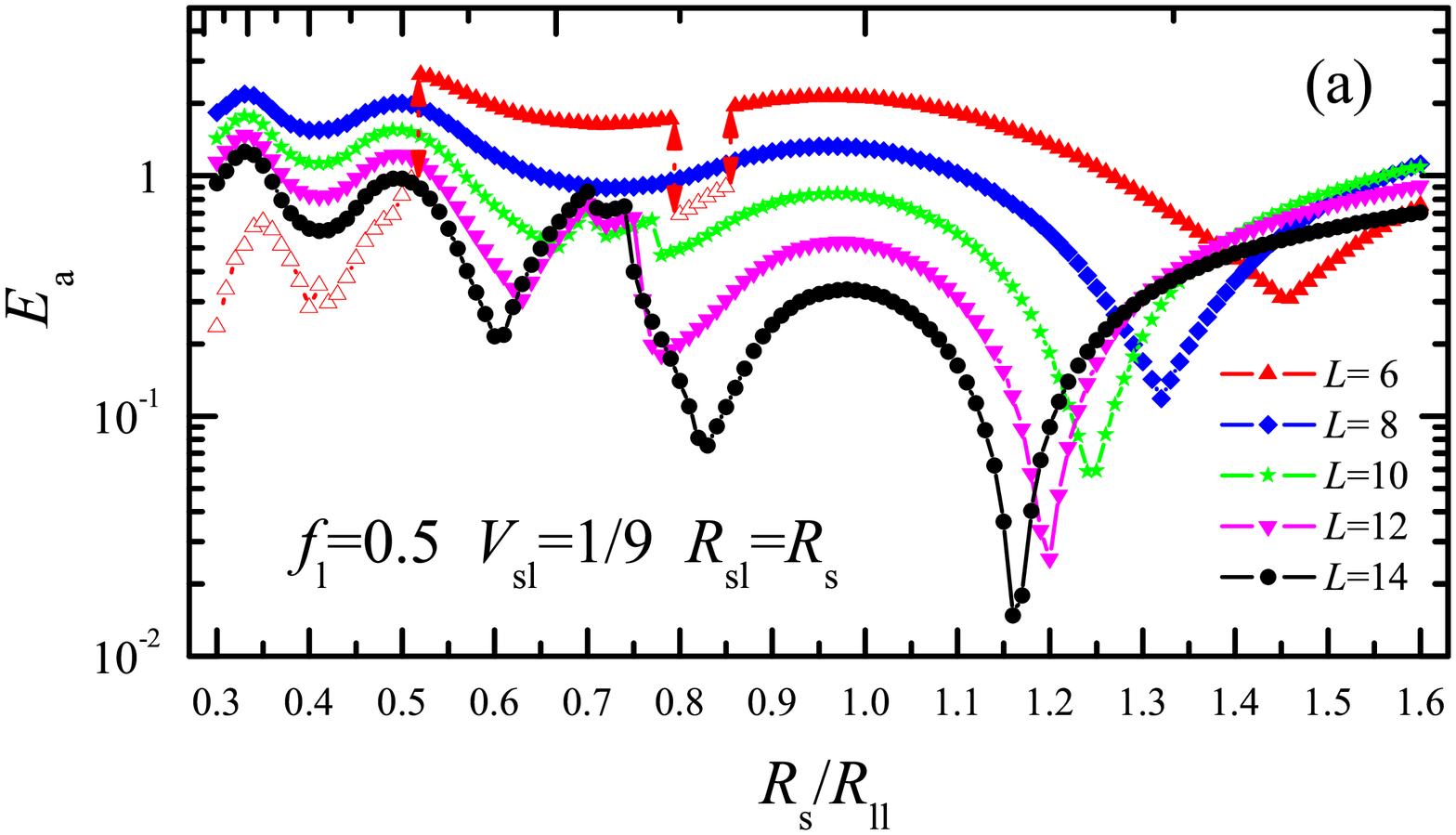} 
\includegraphics[width=10cm]{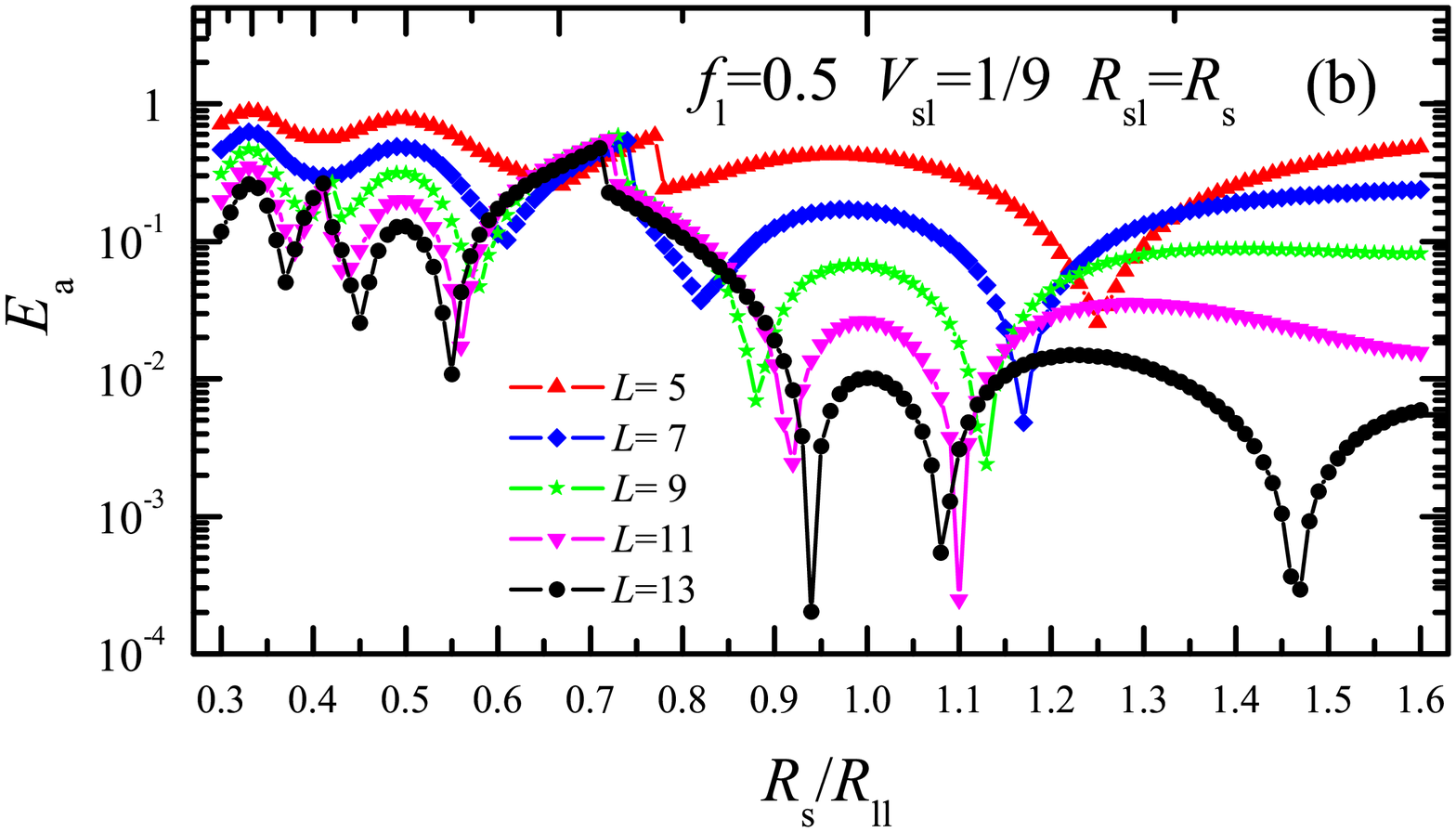} 
\end{center}
\caption{\label{R07} 
Rigid molecule activation energy $E_a$ versus $R_s /R_{ll}$ for
$f_l =0.5$ and $V_{sl}=1/9$.
Unlike figure~\ref{R04c}, here $R_{sl}=R_s$.
(a) is for even $L=6$, 8, 10, 12 and 14; (b) is for odd $L=5$, 7, 9, 11 and 13.
Empty triangles in $L$=6 indicate sliding motion intervals.}
\end{figure}

Varying $R_s$ in figure~\ref{R04c}, we kept fixed
the equilibrium distance $R_{sl}$ for the lubricant-substrate interaction.
More realistically, it might be reasonable to set, e.g., $R_{sl}=R_s$, in which
case, as we observed, the interval of $R_s$ values where rolling prevails
is wider than for fixed $R_{sl}$.
Further preference
for rolling over sliding is found for increasing load $f_l$
and for decreasing interaction strength $V_{sl}$. 
We also note that when sliding wins over 
rolling for $R_s < R_{ll}$,
it 
provides a lower activation energy.
Recalling that $\phi_0=2\pi /L$, the region of parameters
for rolling should increase with $L$ -- a rounder wheel rolls better. 
The $R_s$ dependence of $E_a (R_s)$ for increasing size $L$ (figure~\ref{R07})
shows rolling for all $R_s$ and for all $L \geq 5$,
except for $L=6$ which shows both rolling and sliding (see open symbols in figure~\ref{R07}a).
As $R_s$ varies, the value of $E_a$ changes by more than two orders of magnitude
for even $L$ and more than three for odd $L$, with deep sharp minima
separated by broad maxima. Clearly, by suitably choosing
$R_s /R_{ll}$ a very strong decrease of rolling friction is attainable.

\begin{figure} 
\begin{center}
\includegraphics[width=10cm]{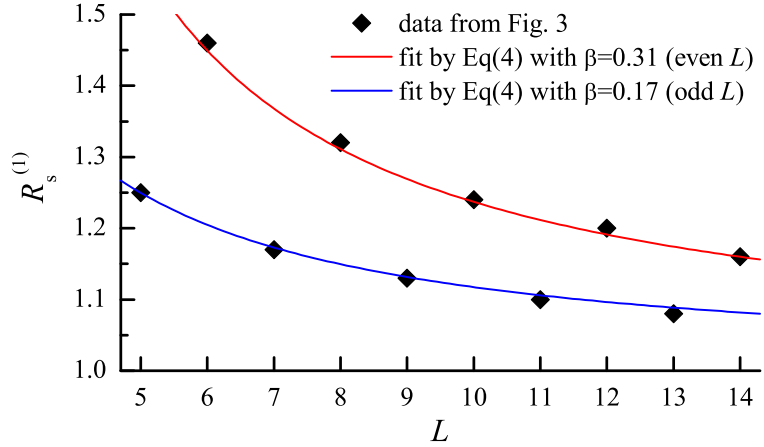} 
\end{center}
\caption{\label{R15} 
Position $R_s^{(1)}$ of the main minimum of $E_a (R_s)$, extracted
from figure~\ref{R07}, for increasing molecular size $L$. Curves are fits to
the cogwheel model~(\ref{Rsmatch}). The asymptotic limit of 1
is still relatively far for reasonable molecular radii. }
\end{figure}

The deep minima of $E_a (R_s)$ are explained by simple engineering -- a ``cogwheel model''.
Consider the molecule as a cogwheel with $L$ cogs, primitive radius $R_m$
and external radius $R^* = R_m +h$, where $h \propto R_{sl}$.
The chord distance between nearest cogs is $R_{ll}^* = 2R^* \sin (\pi/L)$.
Best rolling conditions are expected when $R_{ll}^*$ matches the
substrate potential period $R_s$, i.e., for $R_s^{(1)} = R_{ll}^*$ and its fractions,
$R_s^{(2)} = R_{ll}^*/2$, $R_s^{(3)} = R_{ll}^*/3$, etc.
The main minimum of $E_a (R_s)$ is expected at
\begin{equation}
R_s^{(1)} / R_{ll} = 1+(2h/R_{ll}) \sin (\pi/L).
\label{Rsmatch}
\end{equation}
As shown in figure~\ref{R15}, the cogwheel model (\ref{Rsmatch}) with
$h=\beta R_{sl}$, where 
$\beta$ is a parameter, can fit very well
the shift of minimum position with molecular size $L$. It can
explain its variation with  load (the radius $R^*$ and therefore $h$
decrease as the load grows) as well as with the lubricant-substrate
interaction $V_{sl}$ ($R^*$ and $h$ decrease with $V_{sl}$). 
It also accounts for the even-odd effect
since odd $L$ involves ingraining perfectly one substrate
at a time, justifying why roughly double values of $\beta$
are needed for even relative to odd~$L$.

\section{MD simulation}
\label{simulation}

\begin{figure}[h] 
\begin{center}
\includegraphics[clip,width=10cm]{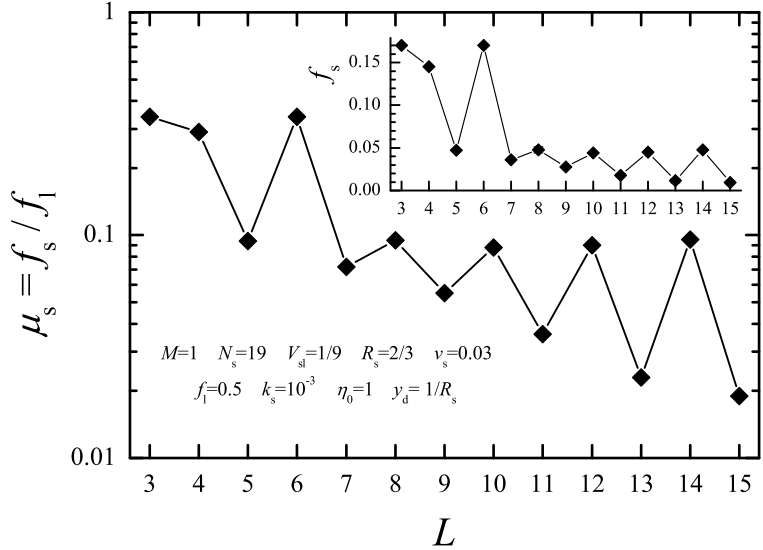} 
\end{center}
\caption{\label{R01}
The static friction coefficient $\mu_s = f_s /f_l$ and
the friction force $f_s$ (inset) for a single circular molecule
as a function of size $L$.
The parameters are
$V_{sl} =1/9$, $R_{sl} =R_s$, $f_l =0.5$, $N_s =19$, $R_s = 2/3$,
$v_s =0.03$, $k_s = 10^{-3}$, $\eta_0 =1$, and $y_d =R_s$.
}
\end{figure}

The simulation results for the static friction of a deformable circular molecule are presented in figure~\ref{R01}.
As one could expect, the $L=3$ or~4 ``circular''  molecule does not 
roll;
instead we observed its ``creep'' with a relatively large friction.
For larger values of $L$, $L \geq 5$, the molecule may either roll or slide.
For rolling in the case of even values of $L$ ($L=6,8,10,12$ and~14) one needs
to break simultaneously two lubricant-substrate bonds
(one connecting the lubricant molecule with the bottom substrate, and one,
with the top substrate).
Therefore, $f_s$ should be approximately independent of $L$,
as indeed is observed in simulation for $L \geq 8$.
For odd values of $L$, $L=5,7,9,11,13$ and~15, 
$f_s$ is at least two times smaller
than for a nearest even $L$ value, 
because one needs to break one bond only at a time.
For all $L \geq 7$ 
the static friction is relatively low, $\mu_s < 0.1$,
and for large odd values the friction may reach quite low values.

\begin{figure} 
\begin{center}
\includegraphics[width=7.5cm]{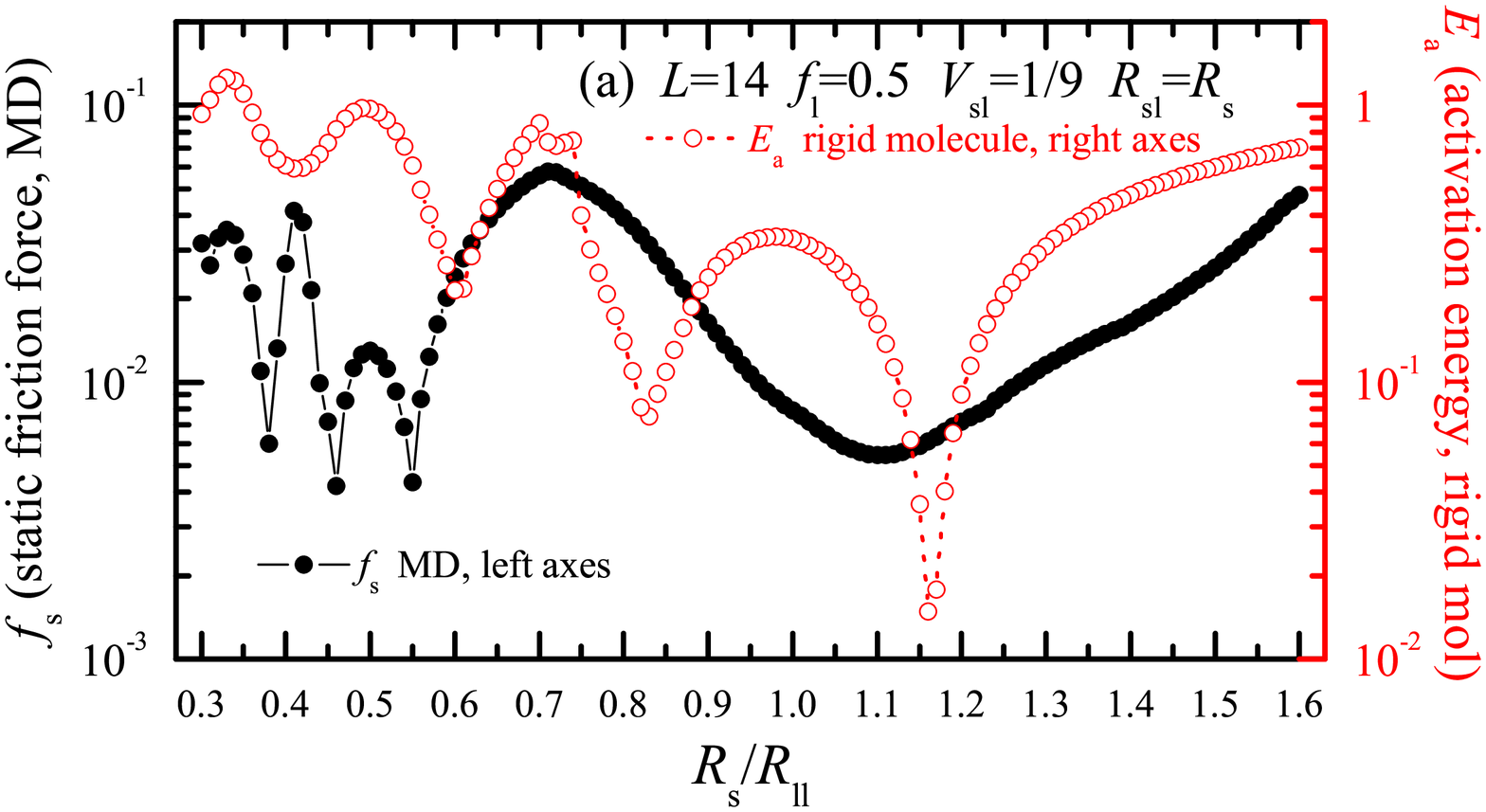} 
\hspace{0.1cm}
\includegraphics[width=7.5cm]{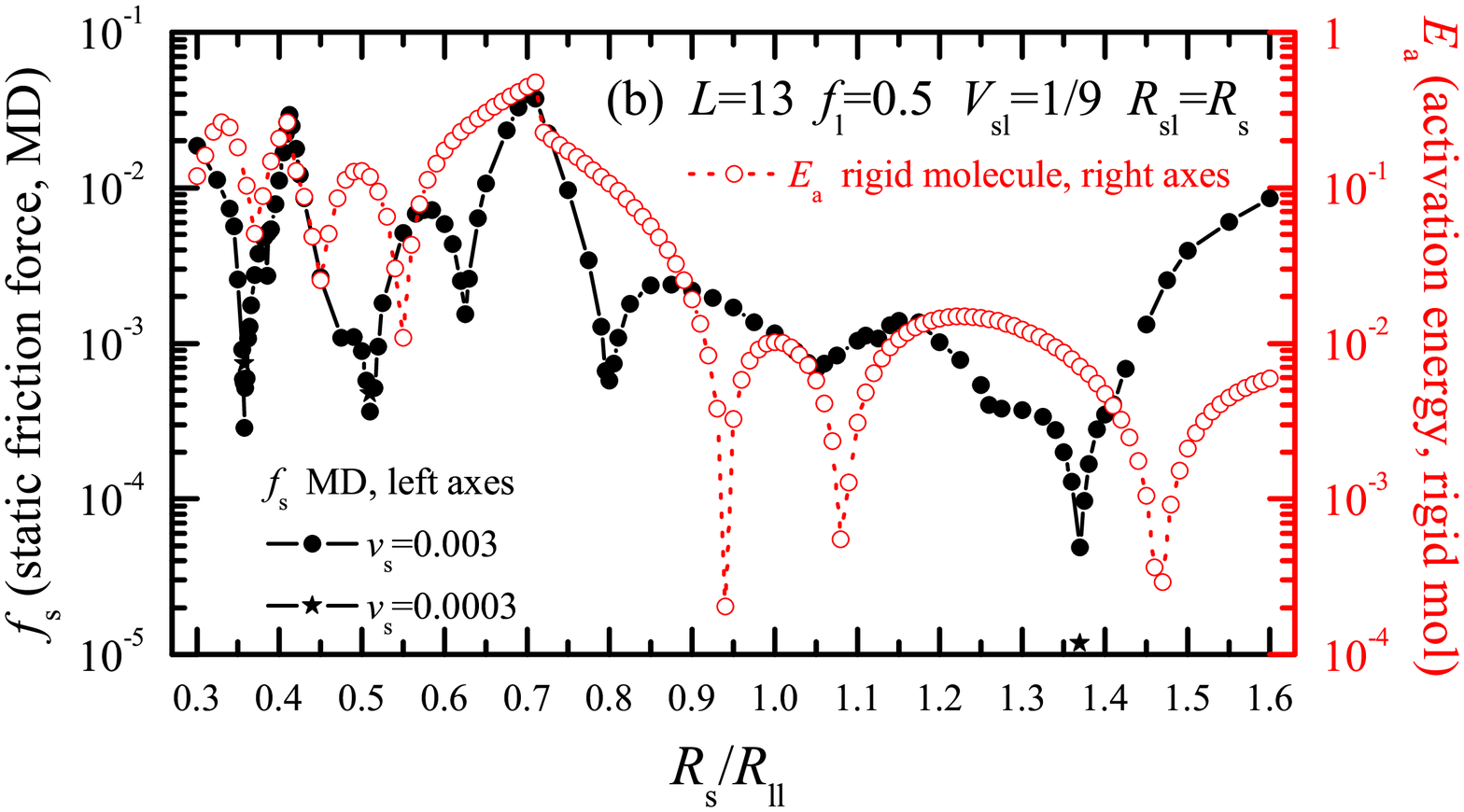} 
\end{center}
\caption{\label{R08} 
The static friction force $f_s$ for the deformable circular molecule
(left axes, $K_m=100$ and $V_{ll}=1$, solid curve and circles for $v_s=0.003$ and stars for $v_s=0.0003$)
and the activation energy $E_a$ for the rigid molecule
(right axes, $K_m=\infty$ and $V_{ll}=\infty$, red open symbols and dotted curve)
as functions of $R_s /R_{ll}$ for $f_l =0.5$, $V_{sl}=1/9$ and $R_{sl}=R_s$.
(a) is for $L=14$, and (b), for $L=13$.}
\end{figure}

\begin{figure} 
\begin{center}
\includegraphics[width=10cm]{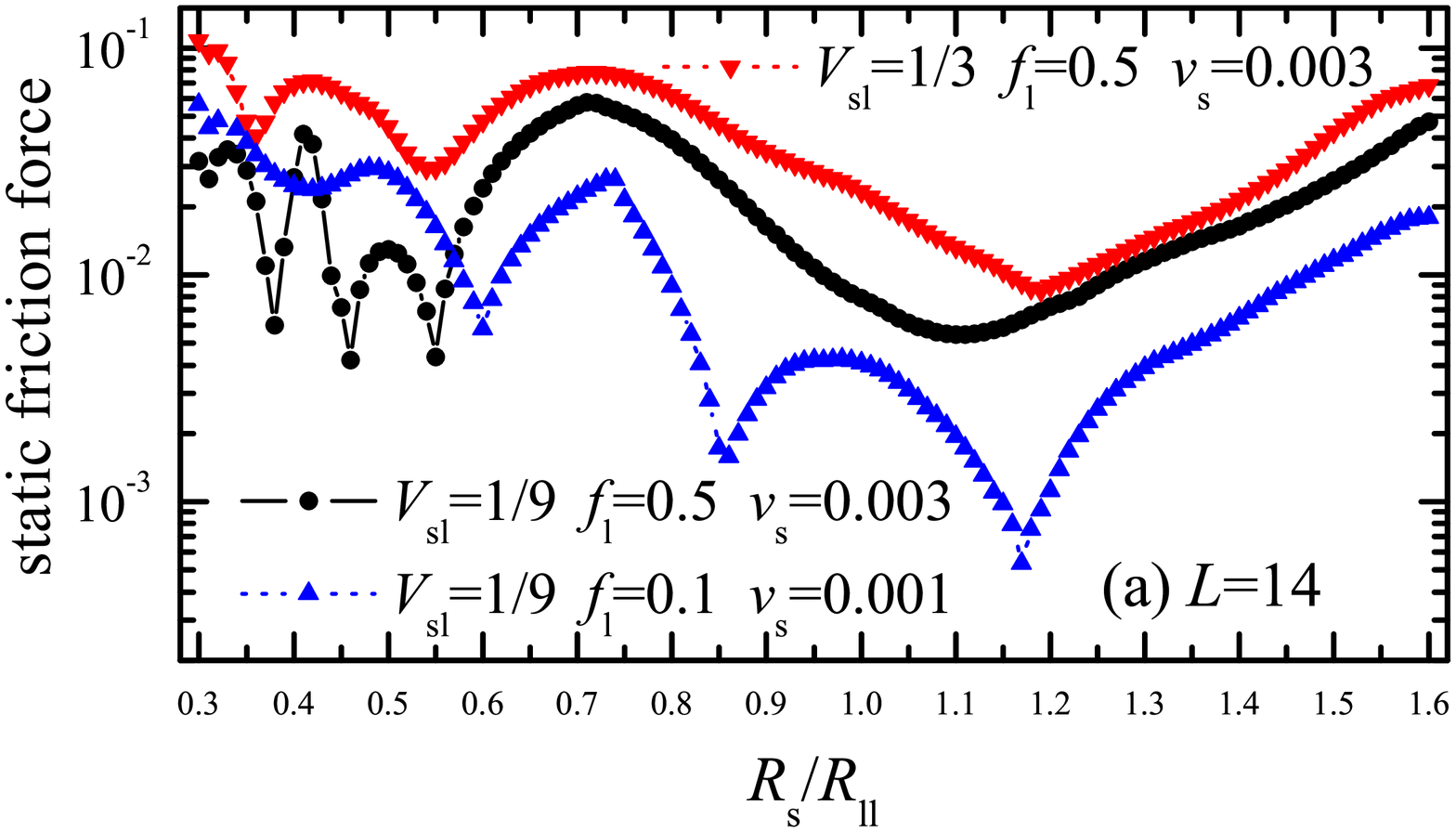} 
\includegraphics[width=10cm]{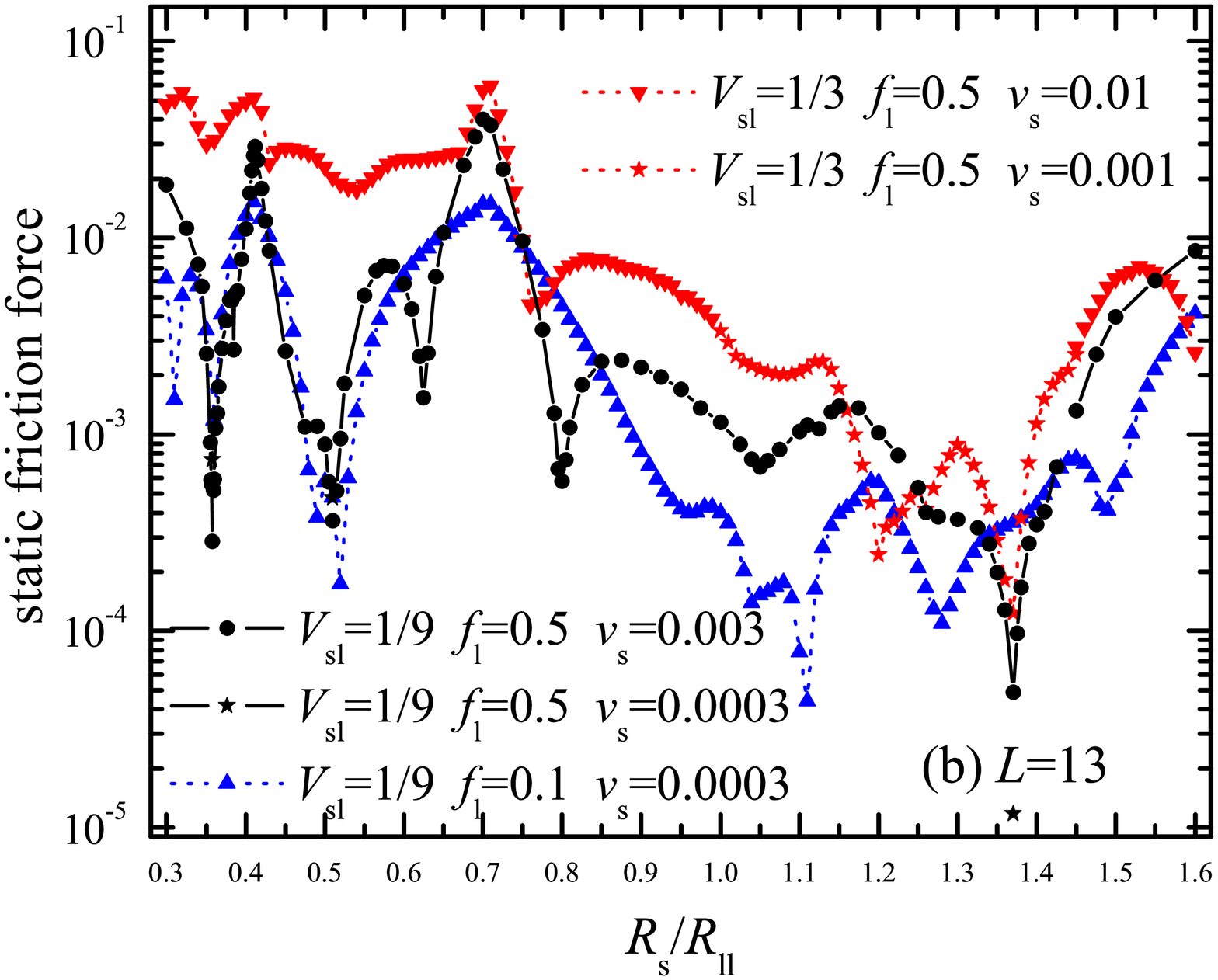} 
\end{center}
\caption{\label{R09} 
The static friction force $f_s$ as a function of the substrate lattice constant $R_s$
for (a) $L=14$ and (b) $L=13$ for different system parameters:
({\it i\/}) $f_l =0.5$ and $V_{sl}=1/9$ (solid curve and circles and stars),
({\it ii\/}) $f_l =0.5$ and $V_{sl}=1/3$ (down triangles and red dotted curve and stars), and
({\it iii\/}) $f_l =0.1$ and $V_{sl}=1/9$ (up triangles and blue dotted curve).
Other parameters are $R_{sl}=R_s$, $K_m=100$, and $V_{ll}=1$.}
\end{figure}

\begin{figure}[h] 
\begin{center}
\includegraphics[width=10cm]{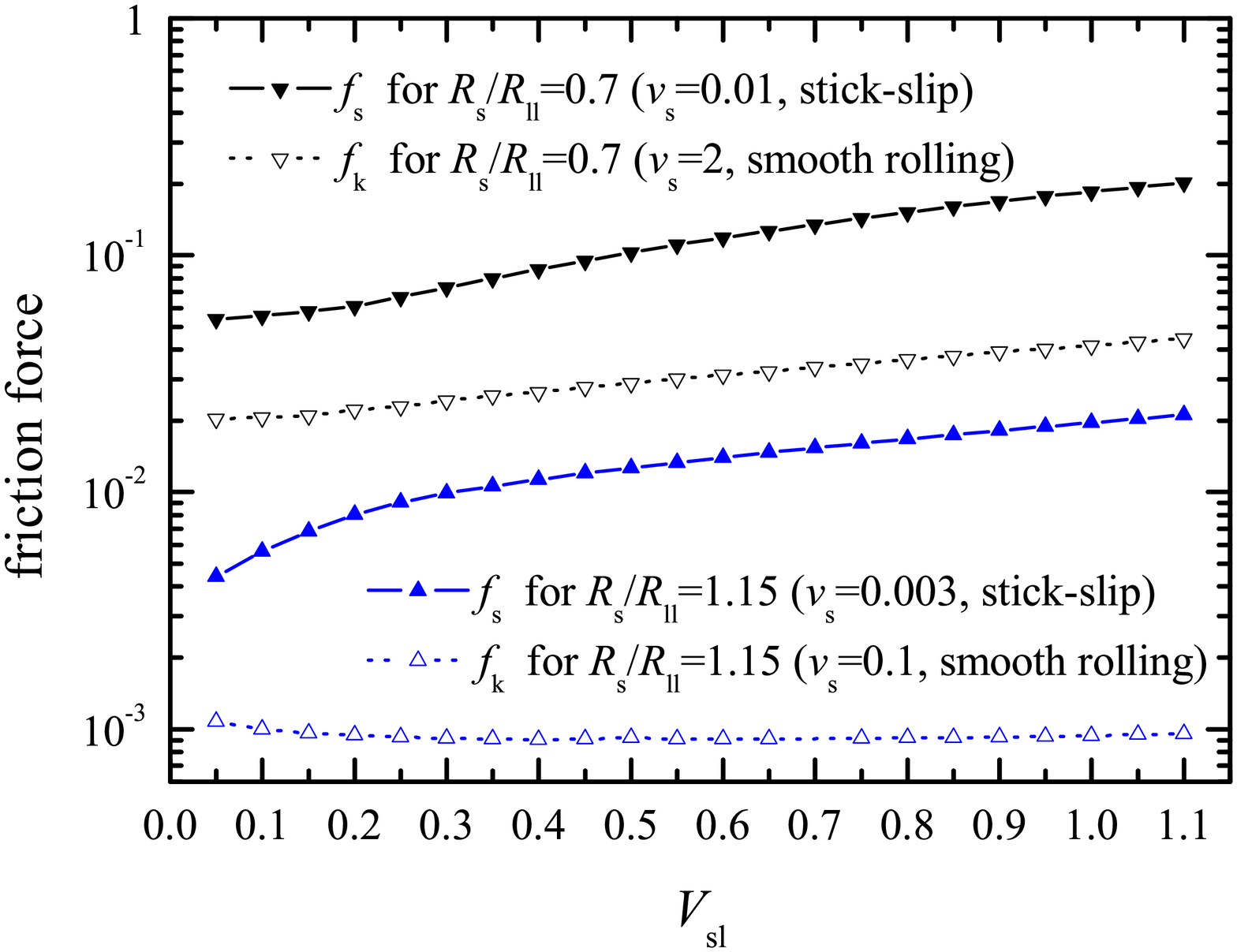} 
\end{center}
\caption{\label{R10} 
The static friction force $f_s$ (solid curves and symbols, stick-slip motion)
and the kinetic friction force $f_k$ (dotted curves and open symbols, smooth rolling)
as functions of the amplitude $V_{sl}$
of lubricant-substrate interaction
for the deformable $L=14$ circular molecule
for two values of the ratio $R_s /R_{ll}=0.7$ (down triangles)
and $R_s /R_{ll}=1.15$ (blue up triangles).
The parameters are $f_l =0.5$, $N_s=19$, $k_s=10^{-3}$,
$\eta_0=1$, $y_d=R_s$, $R_{sl}=R_s$, $K_m=100$, and $V_{ll}=1$.}
\end{figure}

\begin{figure}[h] 
\begin{center}
\includegraphics[width=10cm]{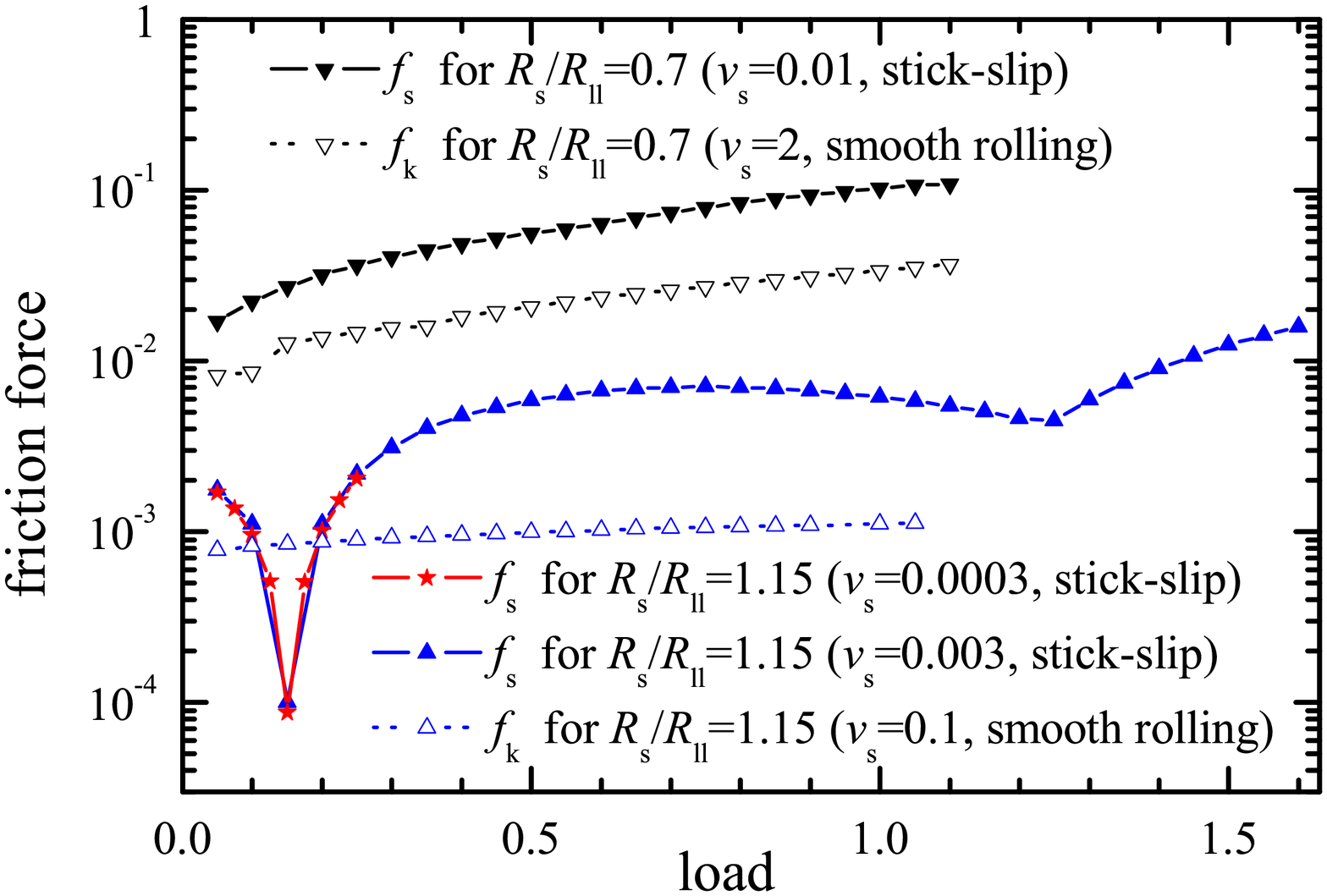} 
\end{center}
\caption{\label{R11} 
The static friction force $f_s$ (solid curves and symbols, stick-slip motion)
and the kinetic friction force $f_k$ (dotted curves and open symbols, smooth rolling)
as functions of the load $f_l$ for the deformable $L=14$ circular molecule
for two values of the ratio $R_s /R_{ll}=0.7$ (down triangles)
and $R_s /R_{ll}=1.15$ (blue up triangles and red stars).
The parameters are $V_{sl}=1/9$, $N_s=19$, $k_s=10^{-3}$,
$\eta_0=1$, $y_d=R_s$, $R_{sl}=R_s$, $K_m=100$, and $V_{ll}=1$.}
\end{figure}

The results obtained for the rigid molecules in section~\ref{rigid}, are qualitatively confirmed
by the static friction force obtained from simulation with deformable molecules.
Figure~\ref{R08} compares the results obtained for the rigid molecule
with the MD calculation of the static friction force of the deformable molecule.
The agreement between these two dependences is reasonable, at least qualitatively.
The friction coefficient $\mu_s = f_s /f_l$ ranges
from $\mu_s \sim 0.1$ at $R_s /R_{ll} \sim 0.7$
to $\mu_s \sim 0.01$ or even $\mu_s \sim 0.001$ at $R_s /R_{ll} \sim 1.1$.
These results are robust to a change of model parameters.
For example, figure~\ref{R09} compares the dependences $f_s (R_s)$ for two values
of the amplitude of lubricant-substrate interaction,
$V_{sl}=1/9$ and~$1/3$, and for two values of the load,
$f_l =0.5$ and~$0.1$.
The next two figures show the dependence of
the static and kinetic friction on $V_{sl}$ (figure~\ref{R10})
and on the load (figure~\ref{R11});
the latter demonstrates that the friction force approximately follows the Amontons law
\begin{equation}
f_{s,k} \approx f_{0s,0k} + \mu_{s,k} f_l.
\label{fskfl}
\end{equation}
Visualization of MD trajectories shows
that for $R_s /R_{ll}=0.7$, where friction is high,
rolling rotation is accompanied by a molecular shift/sliding,
-- much as cogwheels with excessive clearance would do --
while for $R_s /R_{ll}=1.1$, where friction is low,
the motion corresponds to pure rotation, corresponding to
optimal cogwheel coupling.

Simulations showed that the results presented above remain valid at nonzero temperature $T$.
When $T$ increases, we observed both static and kinetic friction force to decrease,
the stick-slip changing to creep and finally to smooth motion at a high temperature. 
\begin{figure}[h] 
\begin{center}
\includegraphics[width=10cm]{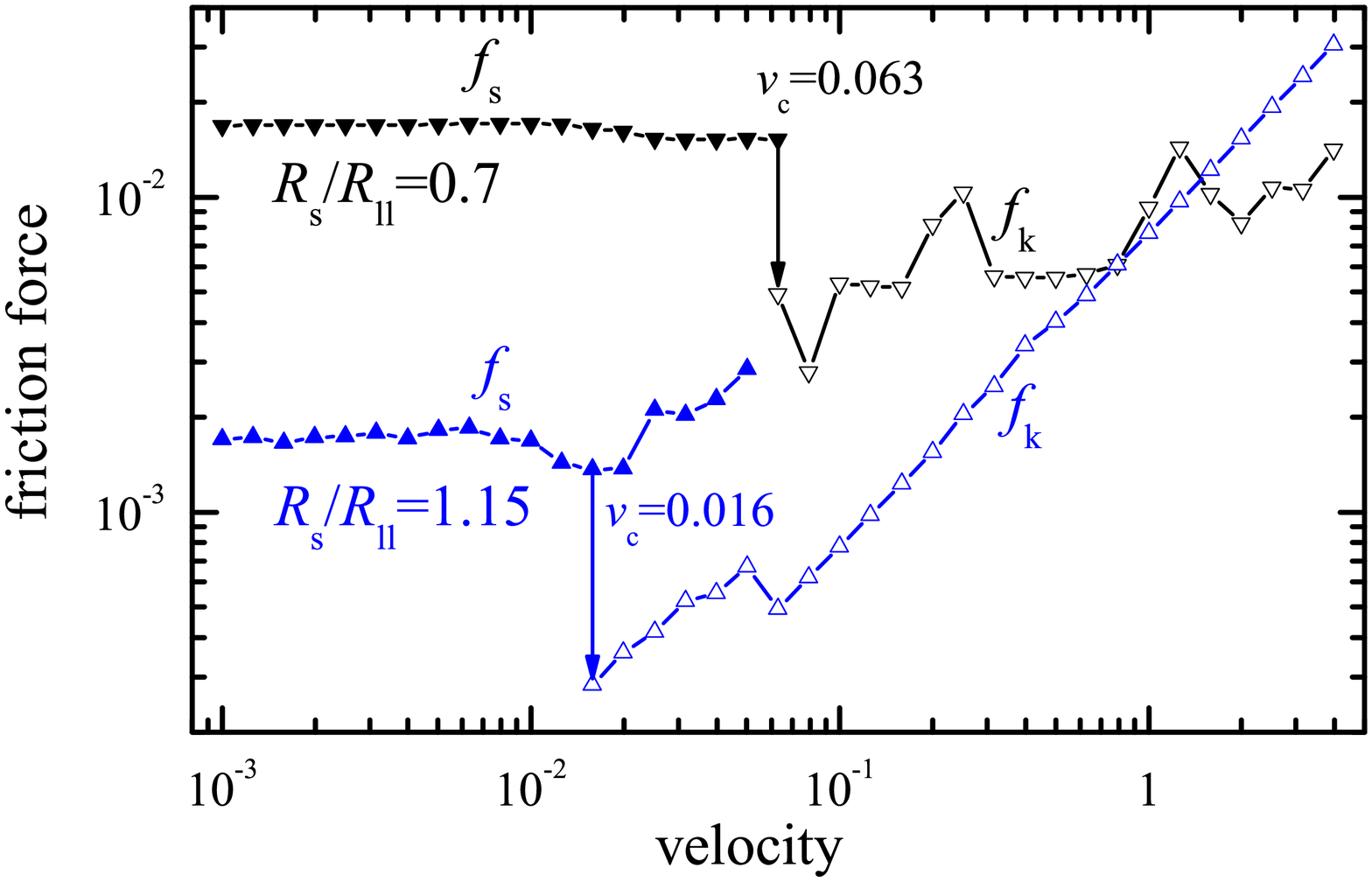} 
\end{center}
\caption{\label{R13} 
The friction force as a function of the driving velocity $v_s$
for two values of the ratio $R_s /R_{ll}=0.7$ (down triangles)
and $R_s /R_{ll}=1.1$ (blue up triangles).
Solid symbols are for the static friction,
open symbols correspond to the kinetic friction;
$v_c$ is the critical velocity of the transition from stick-slip to smooth rolling.
The parameters are $L=14$, $f_l=0.5$, $V_{sl}=1/9$, $N_s=19$, $k_s=10^{-3}$,
$\eta_0=1$, $y_d=R_s$, $R_{sl}=R_s$, $K_m=100$, and $V_{ll}=1$.}
\end{figure}
Moreover, we found a transition from stick-slip to smooth rolling
for increasing velocity (figure~\ref{R13}). The cogwheel effect remains, and for example
calculated static friction for 
$R_s /R_{ll}=0.7$ and $R_s /R_{ll}=1.1$ still differ by a factor of 10 or more.
The critical velocity $v_c$ of the transition from stick-slip to smooth rolling
also differs by a factor of about four in the two cases. Moreover we always find
$f_k \ll f_s$.

The present approach to the single rolling molecule can be extended
to a finite coverage of lubricant molecules.
\begin{figure}[h] 
\begin{center}
\includegraphics[width=10cm]{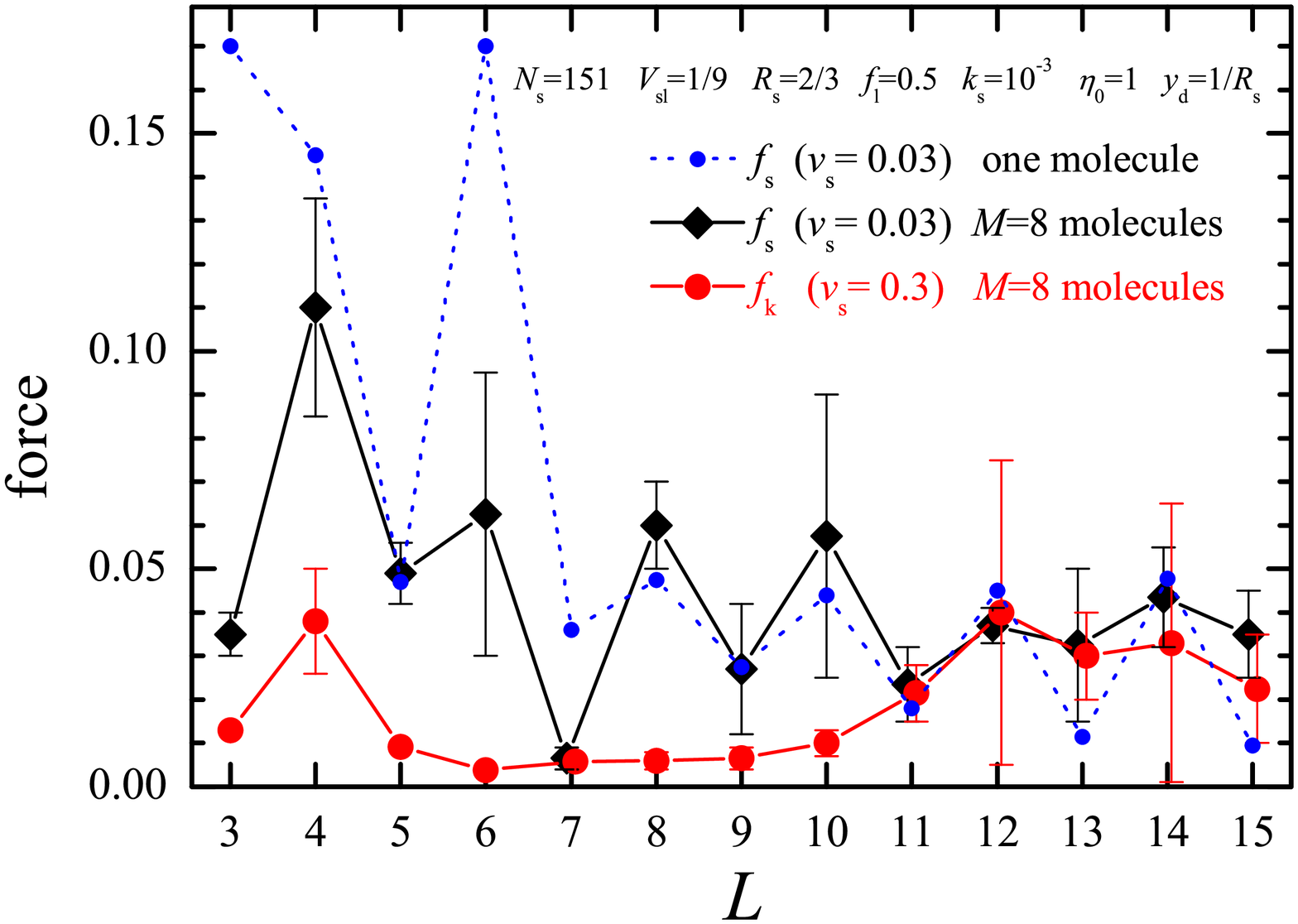} 
\end{center}
\caption{\label{R03} 
The static (diamonds) and kinetic (red circles) friction force
for a finite concentration of circular molecules.
The parameters are $N_s =151$, $f_l=0.5$, $M=8$,
$v_s =0.03$ for the static friction (stick-slip) and
$v_s =0.3$ for the kinetic friction (smooth motion);
other parameters are as in figure~\ref{R01}.
Small blue circles and dotted curve show the results for the single molecule from figure~\ref{R01}.}
\end{figure}
For example, figure~\ref{R03} shows the friction force for a finite concentration
of lubricant molecules, which may be compared with those of figure~\ref{R01} for a single molecule.
These results are for approximately the same load per one lubricant molecule
($f_l N_s /M \approx 9.5$ in both cases), and we used a relatively low
concentration of lubricant molecules, $M/N_s \approx 0.05$, to avoid jams.
The dependence $f_s (L)$ in figure~\ref{R03} is essentially similar to that of figure~\ref{R01},
although the even-odd oscillation of $f_s$ with $L$ are less pronounced at
the finite concentration because of collisions between the molecules.
As for kinetic friction at high driving velocity $v_s=0.3$
for smooth motion, it demonstrates a more monotonic behavior with $L$
without even-odd oscillations.
The function $f_k (L)$ reaches a minimum at $L=6$ where $\mu_k < 0.01$,
and then increases until $L=12$;
at higher values of $L$ the dependence $f_k (L)$ approximately repeats the behavior of $f_s (L)$.

\begin{figure}[h] 
\begin{center}
\includegraphics[width=12cm]{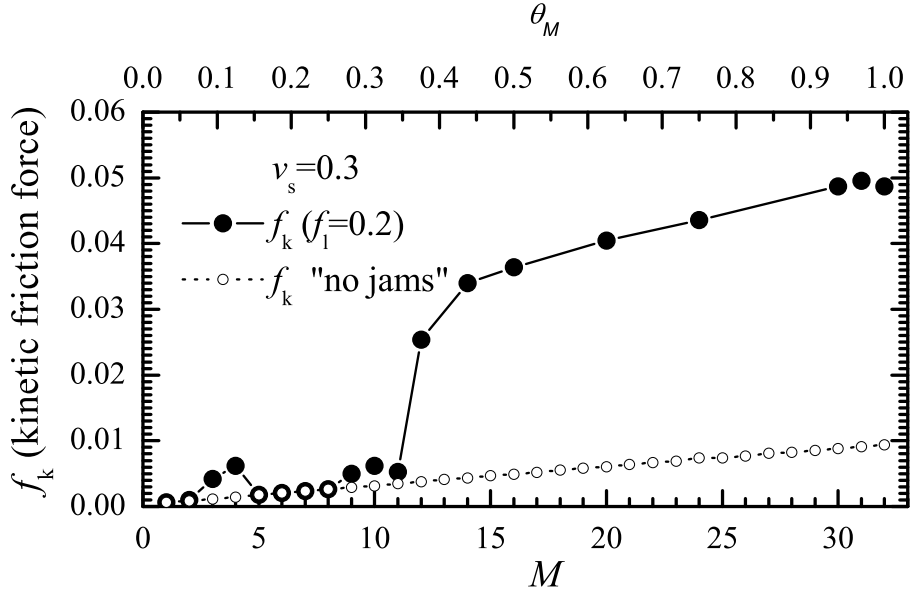} 
\end{center}
\caption{\label{R14}
Dependence of the kinetic friction force at $v_s =0.3$
on the number of circular lubricant molecules $M$ (bottom axes)
or on the dimensionless coverage $\theta_M = M/M_1$ (top axes).
The parameters are:
$L=14$, $N_s =151$, $f_l =0.2$, $R_{s}/R_{ll} =1.15$, $V_{sl} =1/9$, $R_{sl} =R_s$,
$k_s = 10^{-3}$, $\eta_0 =1$, $y_d = R_s$, $K_m =100$, and $T=0.05$.}
\end{figure}
Finally, figure~\ref{R14} shows the dependence of the friction force
on the concentration of lubricant molecules
for $L=14$ and $R_{s}/R_{ll} =1.15$,
which provided a low friction in the single molecule case (figure~\ref{R09}a).
When $M$ increases, the total loading force $F_l =f_l N_s$
is split over the $M$ molecules, so that for a given molecule the load is $f_{l1}=f_l /M$.
As the load decreases with $M$, the friction force per molecule $f_{s1,k1}$ should
also decrease according to~(\ref{fskfl}).
At the same time, the total friction force should increase, $f_{s,k}=M f_{s1,k1}$.
A combined effect is a slow increase of the friction with $M$ as shown in figure~\ref{R14}
with dotted curve and open symbols.
In a real situation, coalescence may lead
to jamming, with molecules blocking their mutual rotation~\cite{B2005}.
In our model, jamming starts already at $\theta_M \approx 0.1$
and completely destroys rolling at $\theta_M > 0.3$
(here $\theta_M =M/M_1$ is the coverage, with $M_1$ the number of
molecules in the monolayer).

\section{Conclusion}

Summarizing, we can extract from our 2D model the following conclusions.
Rolling spherical lubricant molecules can indeed provide
better tribological parameters than sliding atomic lubricants.
The effect may be as large as in macroscopic friction,
where rolling reduces friction by a factor of $10^2 - 10^3$, however
\textit{only for sufficiently low coverage of lubricant molecules,
and for specially chosen values of the ratio}
$R_{s}/R_{ll}$,
corresponding to perfect cogwheel rolling.
To check experimentally these predictions,
it would be interesting to study friction coefficient
for different spherical molecules, different coverages, and different substrates.
Also, the relative
ingraining between the rolling molecule and the substrate may be improved by adjusting
the applied load,
as it was demonstrated experimentally for
the molecular rack-and-pinion device \cite{CGRSGJM2007}.
Inert nonmetal surface (such as perhaps self-assembled monolayers) may represent
a better choice of substrate than metals for fullerenes deposition.
Because 3D rolling has an azimuthal degree of freedom, the
cogwheel effect described should be direction dependent, and rolling
friction should exhibit anisotropy depending on direction.

\ack
This research was supported in part by MIUR PRIN/Cofin Contract No. 2006022847
and by the  Central European Initiative (CEI) whose contribution is gratefully
acknowledged.


\end{document}